\documentclass[reprint, showkeys, showpacs, nofootinbib]{revtex4-1}
\usepackage{lineno}
\usepackage{epsfig}
\usepackage{psfrag}
\usepackage{amsmath, amssymb}
\usepackage{multirow}
\usepackage{graphicx}
\usepackage{wrapfig}

\usepackage{supertabular}

\newcommand{\fil}{\sqrt{k} l}

\usepackage{here}
\begin{document}

\date{\today}
\title{Complex formalism of the linear beam dynamics}
\author{J. Lucas} 
\email{julio.lucas@elytt.com}
\affiliation{ELYTT Energy S.L., Calle de Orense 11, 28046 Madrid}

\author{V. Etxebarria}
\email{victor.etxebarria@ehu.eus}
\affiliation{Depto. de Electricidad y Electr\'onica, Universidad del Pa\'{\i}s Vasco - UPV/EHU, 48940 Leioa, Spain}

\begin{abstract}
It has long been known that the ellipse normally used to model the phase space extension of a beam in linear dynamics may be represented by a complex number which can be interpreted similarly to a complex impedance in electrical circuits, so that classical electrical methods might be used for the design of such beam transport lines.  However, this method has never been fully developed, and only the transport transformation of single particular elements, like drift spaces or quadrupoles, has been presented in the past.  In this paper, we complete the complex formalism of linear beam dynamics by obtaining a general differential equation and solving it, to show that the general transformation of a linear beam line is a complex Moebius transformation. This result opens the possibility of studying the effect of the beam line on complete regions of the complex plane and not only on a single point.  
Taking advantage of this capability of the formalism, we also obtain an important result in the theory of the transport through a periodic line, proving that the invariant points of the transformation are only a special case of a more general structure of the solution, which are the invariant circles of the one-period transformation.  Among other advantages, this provides a new description of the betatron functions beating in case of a mismatched injection in a circular accelerator.
\end{abstract}


\maketitle
\section{Introduction}
Complex variable methods have been widely used in the past to conveniently represent two-dimensional fields and phenomena by essentially identifying the components of a planar vector with the real and imaginary parts of a complex number. This approach has been applied to most branches of science, but particularly it is worth mentioning that in the very early days of electromagnetic theory, even Maxwell included complex variable methods in a chapter of one of his major works to set up potential functions satisfying Laplace's equations in two dimensions \cite{maxwell1881}. Since then, many complex variable formulations elegantly describing a wide variety of electromagnetic phenomena have been proposed. Interestingly enough, this complex formulation has been repeatedly applied in the past  to describe different electromagnetic properties and systems in relation to dynamics, transport and optics of charged particle beams, including electrostatic space-charge phenomena \cite{walker1950}, \cite{kirstein1958}, two-dimensional magnetic fields \cite{beth1966}, \cite{beth1969} or phase plane linear beam optics \cite{Hereward59}. However, the idea was never very popular, as far as beamline dynamics design is concerned, in comparison with transport matrices or Twiss parameters, probably because of the lack of a complete complex formalism which could boost the method to its full potential as applied to this field.

In this paper we extend and develop the idea originally proposed by Hereward \cite{Hereward59} of using a complex number to represent the phase space ellipse describing the motion in a linear beamline. Instead of just using this formulation to treat beam transport lines as electrical circuits by analogy, as it was intended in principle, here the general differential equation governing the complex representation of the phase plane ellipse is obtained and solved. As a result, it is demonstrated that the general transformation of a linear beam line is a subgroup of the well-known complex Moebius transformation. Apart from the elegance and compactness of the general obtained solution in complex formulation, the result contributes to improve our fundamental knowledge of linear beamline dynamics as well as to be able to transform complete regions of the complex phase space instead of single points, as it is the case with the conventional Twiss parameters approach.

As a practical example of use of the new formalism proposed, the method is applied to the important case of beam transport through a periodic line (for instance, through a circular accelerator). By means of the complex formalism we prove the existence of invariant circles under such periodic transformation, which generalizes the classical concept of invariant points of the transformation determined through the Twiss parameters methods, meaning that invariant points in a periodic transport line are indeed invariant circles of null radius. This allows us to reinterpret and more accurately describe, predict, quantify and control important effects measured in practice, such as betatron oscillations and their beating, among other properties.

As described in the following, both in theory and illustrated through example, it becomes clear that the proposed complex formalism gives us a deeper insight on linear beam dynamics\footnote{Note that this paper deals only with linear beam dynamics. However a complex formalism might also be established for non-linear beam dynamics through Birkhoff Normal Forms, which are naturally based on complex formalism}, generalizes classical concepts and has clear implications both in theoretical beam optics and in practical computation and design of present and future beam transport lines.

\section{Definition of the complex parameters}
Although the complex formalism of linear beam dynamics may be obtained without using the Twiss\footnote{In all the following we will be referring to the Courant-Snyder lattice function parameters by using the widespread (though not so proper) term Twiss parameters} parameters, here we will refer to them for the sake of completeness and for easier comparison between the standard and the new proposed formulation.  The phase space ellipse is parametrized by Eq.~(\ref{eq:param}), \cite{Wiedemann93}\cite{Bryant93}, where the Twiss parameters, $\alpha$, $\beta$ and $\gamma$ define the shape of the ellipse, and the emittance $\epsilon$ is related to its area.  In addition, because of the symplecticity of the  dynamics, the relation $\beta \gamma - \alpha^2 = 1$ holds among the ellipse parameters.

\begin{equation}
\gamma x^2 + 2 \alpha x x' + \beta x'^2 = \epsilon
\label{eq:param}
\end{equation}

\begin{figure}[h]
\begin{center}
\psfrag{1}{$-\alpha \sqrt{\epsilon/ \beta}$}
\psfrag{2}{$-\alpha \sqrt{\epsilon/ \gamma}$}
\psfrag{3}{\hspace{6mm}$\sqrt{\epsilon/ \beta}$}
\psfrag{4}{\hspace{5mm}$\sqrt{\epsilon \gamma}$}
\psfrag{5}{$\sqrt{\epsilon \beta}$}
\psfrag{6}{$\sqrt{\epsilon/ \gamma}$}
\psfrag{phi}{$\phi$}
\psfrag{eq1}{$\tan 2 \phi = \frac{2 \alpha}{\gamma-\beta}$}
\epsfig{file=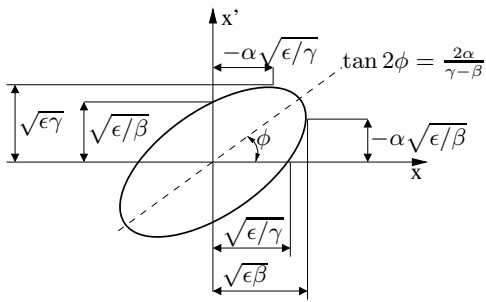, width=6.0cm}
\caption{Twiss parameters defining the phase space ellipse}
\label{fig:elipseDeFase}
\end{center}
\end{figure}

The relationships between the Twiss parameters and the shape of the beam space ellipse may be seen in Fig. \ref{fig:elipseDeFase}.

When the beam is transported through a drift space, the point of maximum divergence will keep its divergence $\gamma_2=\gamma_1=\gamma$, while its position will shift according to $x'$ and the drift length $L$.  This may be stated as:

\begin{equation}
-\alpha_2 \sqrt{\frac{\epsilon}{\gamma_2}}=-\alpha_1 \sqrt{\frac{\epsilon}{\gamma_1}}+ L\,\sqrt{\epsilon \gamma_1}
\end{equation}

So, through a drift space, it holds:

\begin{gather}
\frac{1}{\gamma_2}= \frac{1}{\gamma_1} \\
-\frac{\alpha_2}{\gamma_2}=-\frac{\alpha_1}{\gamma_1}+L
\end{gather}

It can be shown in the same way that through a thin lens of focal length $f$, the following relationships will follow:

\begin{gather}
\frac{1}{\beta_2}= \frac{1}{\beta_1} \\
-\frac{\alpha_2}{\beta_2}=-\frac{\alpha_1}{\beta_1}-\frac{1}{f}
\end{gather}

Now two complex numbers can be defined as:

\begin{gather}
Z=\frac{1}{\gamma}-j \frac{\alpha}{\gamma} \\
Y=\frac{1}{\beta}+j \frac{\alpha}{\beta} 
\end{gather}

It may be seen by direct multiplication that $ZY=1$. In addition, through a drift space we will have, $Z_2=Z_1+j L$, and through a thin lens $Y_2=Y_1+j /f$.

\begin{figure}[h]
\begin{center}
\epsfig{file=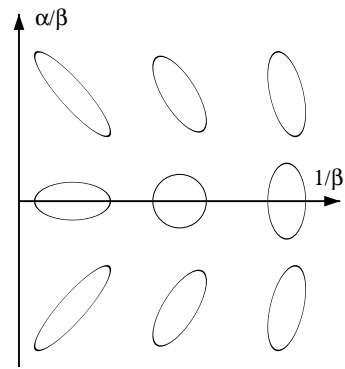, width=4.5cm, bbllx=192, bblly=252, bburx=372, bbury=448, clip=}
\caption{Shape of the phase space ellipses according to their location in the $Y$ plane.  Each ellipse is drawn in a local $x$-$x'$ system.}
\label{fig:elipses}
\end{center}
\end{figure}

If  the $Y$ parameter is used, the size of the beam will be proportional to $\sqrt{\epsilon/\textrm{Re} Y}$, while the lines passing through the origin are of constant $\alpha$.  The upper part of the complex plane corresponds to converging beams and the lower one to diverging ones. Fig. \ref{fig:elipses} shows a qualitative representation of the shape of the phase space ellipses according to their position in the $Y$ plane.

\section{The general differential equation of the complex form}
All the previous material has been studied in the past, starting with the report by Hereward, \cite{Hereward59}, and it is cited extensively, \cite{Regenstreif67}, \cite{Joho80}, \cite{Nemes82}; but as far as it is known to the authors, the general differential equation governing the complex representation of the phase space ellipse has never been obtained. In order to obtain this equation, in the following an infinitesimal displacement through a lens of strength per unit length $k$ will be analyzed \cite{Campos1}.

Because the displacement is infinitesimal, the effect of the lens, which is straightforward in $Y$, can be superposed with the effect of the drift $ds$, which is straightforward in $Z$.  This superposition may be expressed either in terms of $Z$ or $Y$, but the latter representation is preferred, because the real part of $Y$ is related to $\beta$, which, in turn, is related to the beam size:

\begin{multline}
dY=dY_{lens}+dY_{drift}=j k ds+d \left(\frac{1}{Z}\right)_{drift}= \\ j k ds + \frac{-dZ_{drift}}{Z^2}=j k ds - Y^2 j ds = \left( k - Y^2 \right) j ds
\end{multline}

And finally, this results in the following Riccati differential equation:
\begin{equation}
\frac{dY}{ds} + j Y^2 = j k(s)
\label{eq:diff}
\end{equation}

To solve this equation, we start by applying a substitution:
\begin{equation}
Y=-j \frac{u'}{u}
\end{equation}

which converts the Riccati equation into the -well-known to beamline practitioners- Hill's second order linear differential equation:
\begin{equation}
u''+k(s)u=0
\label{eq:kk}
\end{equation}

whose solutions may be expressed, using the superposition principle in linear equations, as a linear combination of the fundamental functions $C(s)$ and $S(s)$, which satisfy $C(0)=1$, $C'(0)=0$, $S(0)=0$, $S'(0)=1$,  \cite{Davis62}. Moreover, two particular solutions of Eq.~(\ref{eq:diff}) can be found, which may be written as:

\begin{gather}
Y_1=-j \frac{C'}{C} \\
Y_2=-j \frac{S'}{S}
\end{gather}

With one particular solution, for instance the one based on $C$, the Riccati equation can be reduced to a first order linear equation using the substitution:

\begin{equation}
Y=Y_1+\frac{1}{z}
\label{eq:YY}
\end{equation}

thus obtaining:

\begin{equation}
z'-2 \frac{C'}{C} z = j
\label{eq:lineal}
\end{equation}

Since a second particular solution of the Riccati equation is already known, a particular solution of Eq.~(\ref{eq:lineal}) can be found as well:

\begin{equation}
z_1=\frac{1}{Y_2-Y_1}= j C S
\end{equation}

where it has been used the fact that $C S'- C' S =1$ because of the constancy of the Wronskian of Eq.~(\ref{eq:kk}).

With this particular solution of Eq.~(\ref{eq:lineal}), its general solution may be obtained as the sum of the general solution of the homogeneous equation and the particular solution:

\begin{equation}
z=A C^2+j C S
\label{eq:generalz}
\end{equation}

where $A$ is an integration constant.  By replacing Eq.~(\ref{eq:generalz}) in Eq.~(\ref{eq:YY}),  the general solution of the Riccati equation is readily obtained:

\begin{equation}
Y=-j \frac{C'}{C}+\frac{1}{A C^2+j C S}
\end{equation}

The integration constant can be found by imposing $Y(0)=Y_0$, to get $A=Y_0^{-1}$.   After some algebra, the final result is obtained:

\begin{equation}
Y=\frac{S' Y_0-j C'}{j S Y_0+C}
\label{eq:moebius1}
\end{equation}

and therefore it is concluded that the transformation of the complex parameter is a Moebius transformation of the shape given by Eq.~(\ref{eq:moebius1}).  A similar expression is found in \cite{Hereward59}, but the proof is restricted to drift spaces and lenses.

We should note now that this result is very important, because the Moebius transformations form group, i.e. the composition of Moebius transformations is a new transformation.  Since any general transformation following Eq.~(\ref{eq:diff}) may be expressed as a composition of individual transformations, it may be concluded that if each single beamline element is described by a Moebius transform, then also the line will be described by a Moebius transform.

\section{Examples of common beam line elements in complex form}
By introducing the well-known transport matrices values of common beam line elements  \cite{Wiedemann93} \cite{Bryant93} in the general result of Eq.~(\ref{eq:moebius1}), some illustrative examples of the Moebius transformations associated to these common elements can be built.

\subsection{A drift space of length $L$}
\begin{equation}
Y_2=\frac{Y_1}{j L Y_1+1}
\end{equation}

\begin{figure}[h]
\begin{center}
\epsfig{file=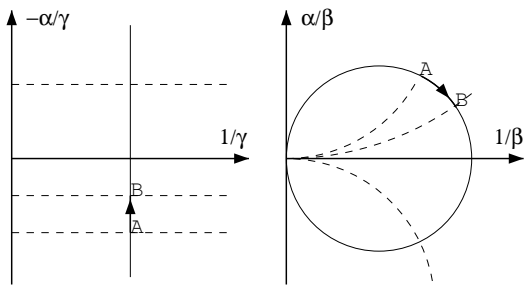, width=7.0cm, bbllx=22, bblly=200, bburx=377, bbury=389, clip=}
\caption{Effect of a drift on the $Z$ and the $Y$ plane}
\label{fig:drift}
\end{center}
\end{figure}

The effect on the complex plane of the drift space in $Z=Y^{-1}$ is a vertical displacement in the upward direction.  The complex inverse of a straight line in the complex plane is a circle passing through the origin.  Therefore the trajectory in the $Y$ plane will be a sector of a circle passing through the origin and tangent to the imaginary axis. The particle will move clockwise, because the inversion implies a change of sign with respect to the movement as seen from the origin.  Fig. \ref{fig:drift}, shows the effect of a drift on the $Z$ and $Y$ planes.

It can also be seen that the length of the drift is measured as the difference in the vertical distance of the two extreme points defining the drift, $A$ and $B$.  If all the horizontal lines of the $Z$ plane were labeled by their constant pure imaginary coordinate $-\alpha/\gamma$, these lines would be transferred to the $Y$ plane as circles passing through the origin and tangent to the real plane.  Each of these circles correspond to a certain $-\alpha/\gamma$.

\subsection{A thin lens of focal length $f$}
\begin{equation}
Y_2=Y_1+\frac{j}{f}
\end{equation}
This represents a vertical displacement upward when the lens is focusing and downward when it is defocusing.

\subsection{A thick lens of strength $k$ and  thickness $l$}

We assume here an ideal horizontally focusing thick quadrupole of length $l$ with constant focusing strength $k(s)=-k$ with $k$ positive.

\begin{equation}
Y_2=\frac{\cos (\fil) Y_1+ j \sqrt{k} \sin (\sqrt{k} l)}{j \sin (\sqrt{k} l)/\sqrt{k} Y_1+\cos (\fil)}
\label{eq:thicklens}
\end{equation}
A similar equation is given in \cite{Lichtenberg69} without proof.

This transformation is, as well, a circle centered in the real axis.  This may be seen by dividing the numerator and denominator of Eq.~(\ref{eq:thicklens}) by $\cos{\sqrt{k} l}$ and considering

\begin{equation}
Y_2=\frac{Y_1+ j \sqrt{k} \tan (\sqrt{k} l)}{j \tan (\sqrt{k} l)/\sqrt{k} Y_1+1}=\frac{Y_1+ j \sqrt{k} u}{j u/\sqrt{k} Y_1+1}
\end{equation}

as a Moebius transform of u=$\tan(\sqrt{k} l)$, ie. the real axis.  As it is known, Moebius transformations convert lines to circles (or lines in some cases). In order to obtain the parameters of the circle, the Riccati equation can be written in terms of its real and imaginary components, $Y=x+j y$.

\begin{align}
\frac{dx}{ds} &= -2 x y \\
\frac{dy}{ds} &= k-\left(x^2-y^2\right) \label{eq:dyds}
\end{align}

It is observed in the above equation that when the sign of $y$ changes, only the horizontal derivative changes its sign. This is the expected behavior from a circle centered on the real axis following the equation:

\begin{equation}
\left( x- x_c \right)^2+y^2=R^2
\end{equation}

where $x_c$ represents the center coordinate on the real axis and $R$ is the radius of the circle.

The circle parameters can be readily obtained by noting that when the vertical derivative Eq.~(\ref{eq:dyds}) cancels, this corresponds to a maximum $y$, i.e., the point with $(x_c, R)$ coordinates.  Thus, the following pair of equations defining the circle passing through the point $\left( x_0, y_0 \right)$ and with a focusing strength of $k$, can be written:

\begin{gather}
\left( x_0- x_c \right)^2+y_0^2=R^2\\
k-x_c^2+R^2=0
\end{gather}

so that the desired parameters will be:
\begin{gather}
x_c=\frac{k+x_0^2+y_0^2}{2 x_0} \\
R=\sqrt{x_c^2-k}
\end{gather}

It is worth to notice here that, when $k>0$, the circle is fully contained in the right part of the complex plane, so that the complex parameter is bounded, while when $k<0$, the circle is partly contained on the left part of the complex plane.

\section{The complex parameters for a circular accelerator}
In a circular accelerator, or in general a periodic line, the beam is expected to repeat its configuration in phase space.  This is equivalent to operate in one of the fixed points\footnote{This section makes use of some mathematical properties of the Moebius transformation, which are summarized in the Appendix for completeness. In particular in the following we will make use of the matrix representation of the Moebius transformation, the fixed points of the transformation and the circle preserving properties, all of which are discussed in the Appendix.} of the Moebius transformation of Eq.~(\ref{eq:moebius1}). Since the fixed point must have a real part, we will reproduce here the classical condition $|C+S'| < 2$ required in order to have a periodic solution for a periodic lattice.  

This result can be expressed in the language of Moebius transforms via the classification of the transformation through the parameter $\sigma$.  This parameter is invariant through any equivalence transformation, and is obtained as:

\begin{equation}
	\sigma=\frac{\left(a+d\right)^2}{ad-bc}-4 = \left( C + S'\right)^2-4
\end{equation}

which is the quotient of the square of the trace by the determinant minus four.  This definition of the transformation invariant is made so as to have a zero value for the identity transformation.  Regarding this parameter, all Moebius transformations may be classified as,

\begin{equation}
	\begin{cases}
		\textrm{Elliptic if } -4 \le \sigma < 0\\
		\textrm{Proper hyperbolic if } \sigma > 0\\
		\textrm{Improper hyperbolic if } \sigma \le -4\\
		\textrm{Loxodromic if } \sigma \textrm{ is not real}
	\end{cases}
\end{equation}

The behavior of an iterative application of the same Moebius transform with respect to the fixed points is fully described by the transformation type. The most important aspect for the transverse beam dynamics study is that only for the elliptic transformation does none of the invariant point represent an attractor.  That is, for all other transformation types, the iterative application of the transformation has one of the invariant points as a limit. In addition, the invariant points for the case $|C+S'| \ge 2$ lie on the imaginary axis and therefore $\beta \rightarrow \infty$, which shows that the beam will grow without limit in size for the hyperbolic case. This condition will ensure as well that only Moebius transformations of the elliptic type lead to periodic solutions in periodic lattices.

\subsection{Structure of the periodic solutions of beam transport in complex form}

The Twiss parameters can readily be expressed as a function of the fundamental solutions:

\begin{gather}
\beta=\frac{1}{\textrm{Re} Y_+}= \frac{2 S}{\sqrt{4-\left( C+S'\right)^2}} \\
\alpha = \frac{\textrm{Im} Y_+}{\textrm{Re} Y_+}=\frac{C-S'}{\sqrt{4-\left( C+S'\right)^2}}
\end{gather}

This is, of course, a classical result of the theory of the Twiss parameters.  We will use now the theory of the complex transformation to obtain the structure of the solutions of the beam transport on a periodic line. Note that these solutions are not easily obtained from the classical theory, and that a beautiful and useful structure can be unveiled for these solutions when seen under the light of the Moebius transformation.

First, we will prove that if there are two circles invariant under the Moebius transformation, a one-dimensional set of circles being also invariant can be found.  A pencil of circles is formed by the linear combination of two circles:

\begin{equation}
	\mathfrak{C}\left(\lambda_1, \lambda_2\right) = \lambda_1 \mathfrak{C}_1+
\lambda_2 \mathfrak{C}_2
\label{eq:pencil1}
\end{equation}

It can be easily proved that if $\mathfrak{C}_1$ and  $\mathfrak{C}_2$ are invariant circles, all the circles of their pencil are invariant.  The pencil is one-dimensional, because Eq.~(\ref{eq:pencil1}) may be multiplied by a constant without modifying the circle of the pencil.

Now we can use the zero radius circles with center in the fixed points of the transformation as the basis of the pencil of invariant circles.  We will call the two fixed points $Y_{+}$ and $Y_{-}$ respectively, according to the sign of the real part of the fixed points given by Eq.~(\ref{eq:fixed}).  Their respective Hermitian matrices are:

\begin{equation}
\mathfrak{C}_\pm = 
\begin{bmatrix}
	1 & -Y_\pm \\
	-\overline{Y}_\pm & Y_\pm \overline{Y_\pm}
\end{bmatrix}
\end{equation}

The parametrization of the invariant circles will be:

\begin{equation}
	\mathfrak{C}_\lambda= \left( 1+\lambda \right)\mathfrak{C}_{+} 
	                               -\lambda \mathfrak{C}_{-} 
\end{equation}

This parametrization ensures that the $A$ term of $\mathfrak{C}_\lambda$ (see Eq.~(\ref{eq:circle3})) is equal to unity, so that the circle is normalized, and that $\lambda=0$ corresponds to the invariant point with positive real part.  It is now possible to obtain the invariant circle of the invariant pencil that passes through any point $y$, by solving $\lambda$ from the equation:

\begin{equation}
	\underline{y}^H \mathfrak{C}_\lambda \underline{y}=0
\end{equation}

The solution of the invariant circle passing through point $y$ will be:

\begin{equation}
	\mathfrak{C}_\lambda=\frac{-\underline{y}^H \mathfrak{C}_{-} \underline{y}}{\underline{y}^H \mathfrak{C}_{+} \underline{y}-\underline{y}^H \mathfrak{C}_{-} \underline{y}} \mathfrak{C}_{+}+\frac{\underline{y}^H \mathfrak{C}_{+} \underline{y}}{\underline{y}^H \mathfrak{C}_{+} \underline{y}-\underline{y}^H \mathfrak{C}_{-} \underline{y}} \mathfrak{C}_{-}
\label{eq:cl}
\end{equation}

The general result of Eq.~(\ref{eq:cl}) can be particularized to the case of the movement around the fixed points given by Eq.~(\ref{eq:fixed}) for the elliptic case.  For this transformation, the two fixed points are symmetric respect to the imaginary axis, i.e. $Y_{-} =-\overline{Y}_{+}$.  If we call $y_0$ the fixed point with positive real part (the one with physical meaning), then the matrices of the two zero radius fixed points will be:

\begin{equation}
\mathfrak{C}_{+} = 
\begin{bmatrix}
	1 & -y_0 \\
	-\overline{y}_0 & y_0 \overline{y_0}
\end{bmatrix} \hspace{1cm}
\mathfrak{C}_{-} = 
\begin{bmatrix}
	1 & \overline{y}_0 \\
	y_0 & y_0 \overline{y_0}
\end{bmatrix}
\end{equation}

Now, by applying Eq.~(\ref{eq:cl}), it is possible to find the invariant circle that passes through a certain point $y$.  The solution is a circle centered at a point $y_c$ given by:
\begin{equation}
y_c=\frac{|y|^2+|y_0|^2-2 \textrm{Im}(y) \, \textrm{Im}\left(y_0\right)}{2 \textrm{Re}\left(y\right)}+ i \textrm{Im}\left(y_0\right)
\end{equation}

meaning that the invariant circles have the center at the same horizontal line as the fixed points. The radius of the invariant circle will then be:

\begin{equation}
R = \frac{|y-y_0| \, |y+\overline{y_0}|}{2 \textrm{Re}\left(y\right)}
\end{equation}

With the knowledge of the invariant circle, it is possible to obtain the maximum beam phase-space limit, which will be given by:

\begin{equation}
\frac{1}{\beta} \ge \frac{|y|^2+|y_0|^2-2 \textrm{Im}(y) \, \textrm{Im}\left(y_0\right)-|y-y_0| \, |y+\overline{y_0}|}{2 \textrm{Re}\left(y\right)}
\end{equation}

This result may be very useful to determine the maximum beam size that may be caused by a mismatch in the injection of the beam into a periodic line.

\subsection{A numerical example}
\begin{figure}[h]
\begin{center}
\epsfig{file=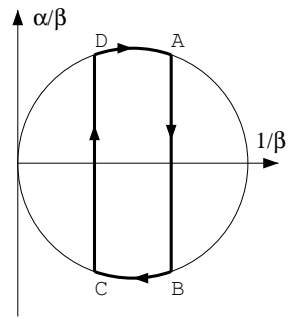, width=4cm, bbllx=-20, bblly=400, bburx=176, bbury=605, clip=}
\caption{Trajectory on the Y plane of a beam through a thin lens FODO cell}
\label{fig:fodo2}
\end{center}
\end{figure}

As an example of the use of the theory of invariant circles on a periodic transport line, let us analyze the structure of the solutions of the beam when injected, not necessarily well matched, on a FODO line.  In a qualitative way, we can show the evolution of the beam along the line for a well matched condition in Fig. \ref{fig:fodo2}.  The focusing thin lens is the line $CD$; the upper circle arc $DA$ is the drift going to the defocusing lens; $AB$ is the defocusing lens and $BC$ is the drift space going to the focusing lens.

It may be worth to notice that the relationship between the Twiss parameters and the complex ones in beam dynamics is similar to that existing between the Bode and the Nyquist plots in control theory. In the first one, the amplitude and the phase are expressed in two different plots, but in exchange the position (frequency) at which the gain is expressed is explicitly indicated in the chart.  In the second one, a complex number provides all information but as a drawback, the position (frequency) at which the value is given, must be provided with a label attached to a certain number of points.

\begin{figure}[H]
	\centering \epsfig{file=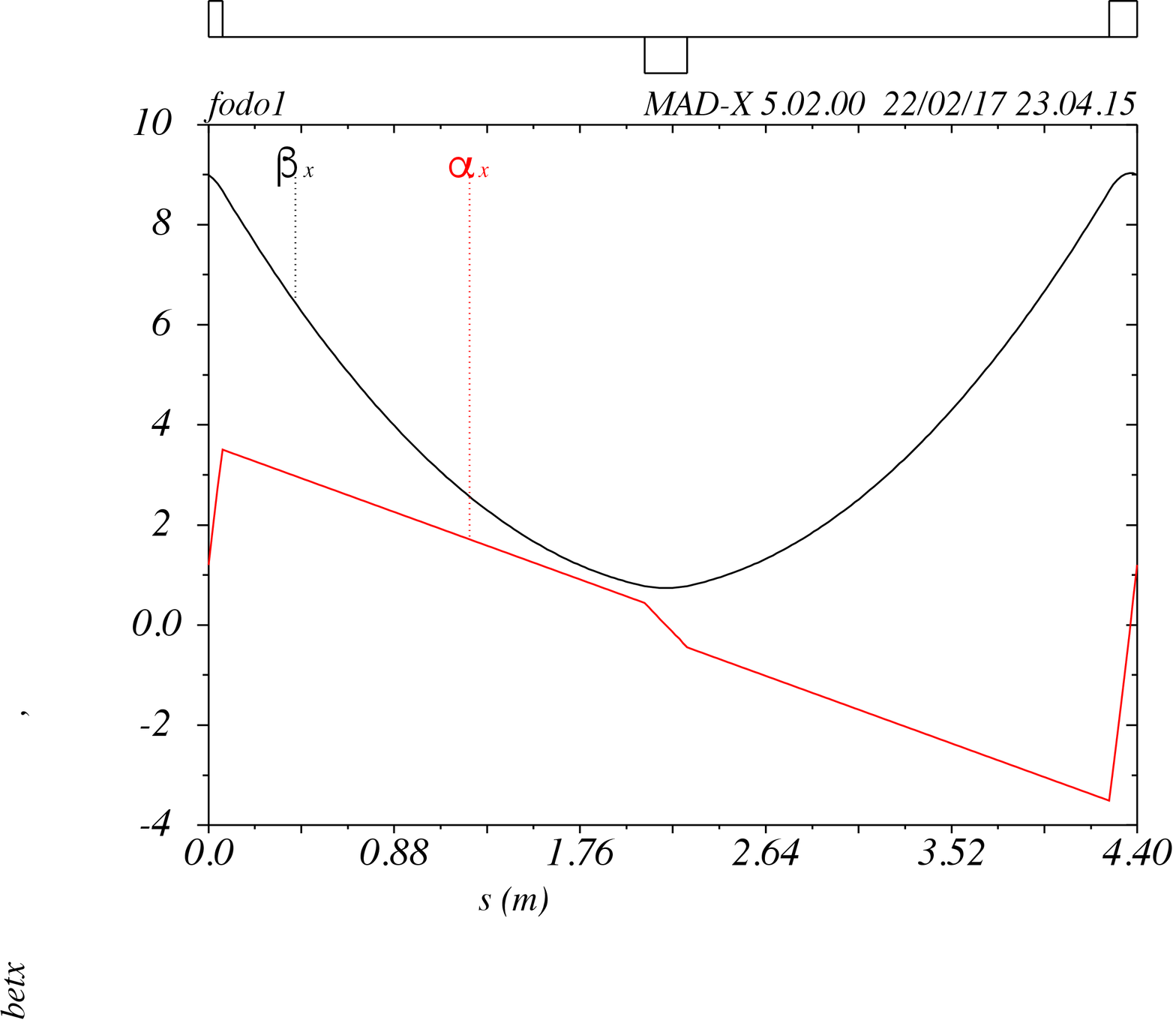, angle=270, width=8cm, bbllx=0, bblly=69, bburx=481, bbury=648, clip=}
	\caption{Horizontal Twiss parameters in the example cell}
	\label{fig:fodoCell}
\end{figure}

To illustrate the complex formulation in a more quantitative situation, next we will define a thick lens FODO cell with quadrupoles of length 0.2~m and drift spaces 2~m long.  The strength of the quadrupoles is $\pm \textrm{4 m}^{-2}$. Arbitrarily, we will analyze the behavior of the horizontal solution at 1/3 of the length of the focusing quadrupole.  The cell, with the horizontal Twiss parameters for the injection matched at the invariant point, may be seen in Fig. \ref{fig:fodoCell}.

\begin{figure}[H]
	\centering \epsfig{file=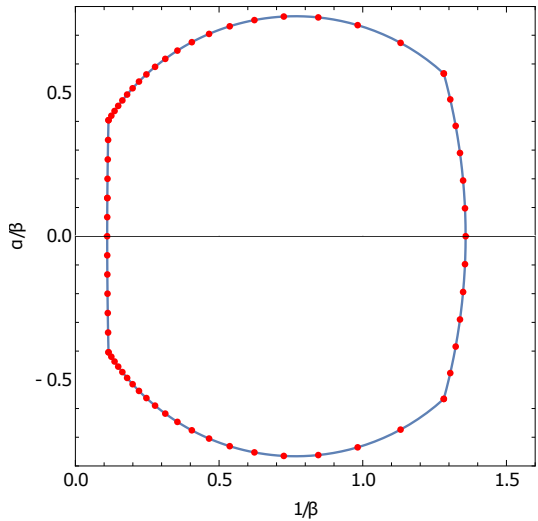, width=7cm}
	\caption{The same cell described in the complex plane}
	\label{fig:fodoCell2}
\end{figure}

The same cell may be described by the movement of the beam point in the complex plane.  This is shown in Fig. \ref{fig:fodoCell2}.  The matching parameters at the injection point are $\beta\approx 9~m$ and $\alpha\approx 1.1$, or the corresponding complex parameter.

\begin{figure}[H]
	\centering \epsfig{file=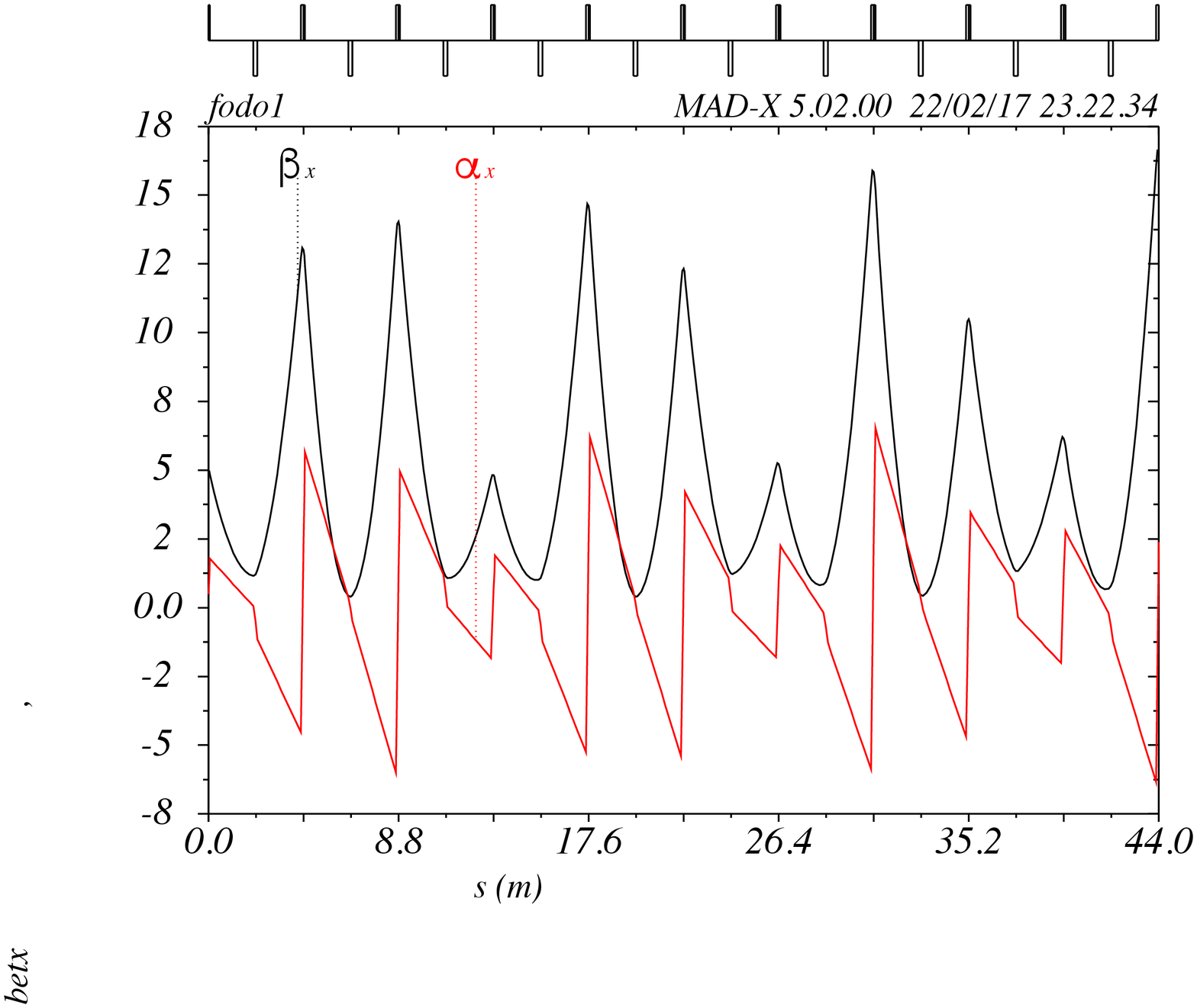, angle=0, width=8cm, bbllx=140, bblly=90, bburx=740, bbury=588, clip=}
	\caption{Horizontal Twiss parameters in the example cell for the case of a non-matched injection.}
	\label{fig:fodoCell3}
\end{figure}

We can now use the theory of invariant circles to study the behavior of the periodic FODO in case of a mismatched injection.  In this case, the beam will oscillate around the periodic parameters in such a way that it will be difficult to find any apparent order. For instance, at Fig. \ref{fig:fodoCell3} the beam has been injected with a $\beta=5~m$ and $\alpha=0.5$. Under this condition, the beam wiggles around the ideal fixed point.  Nevertheless, the classical theory does not provide any information about the maximum amplitude of the Twiss parameters during the oscillation.

\begin{figure}[H]
	\centering \epsfig{file=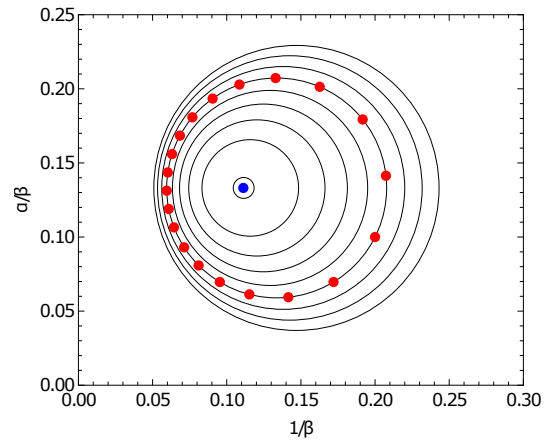, width=7cm}
	\caption{The pencil of invariant circles at 1/3 of the length of the quadrupole.  The position of the beam passes during several periods, as well as the fixed point, are shown}
	\label{fig:fodoCell4}
\end{figure}

The situation becomes much more clear when the different complex points are plotted in the complex plane after each period. This is shown in Fig. \ref{fig:fodoCell4}.  The points must remain in the circle of the invariant pencil of circles that passes through the point defined by the injection parameters.  At some passages, the point will be at the right side of the fixed point, corresponding to a smaller beam and at other passages, the point will be at the left side, which corresponds to a larger beam.  Nevertheless, the beam size will be bounded by the left-most side of the invariant circle, which is easily obtained through the parameters of the $\mathfrak{C}_\lambda$ invariant circle of Eq.~(\ref{eq:cl}).

\begin{figure}[H]
	\centering \epsfig{file=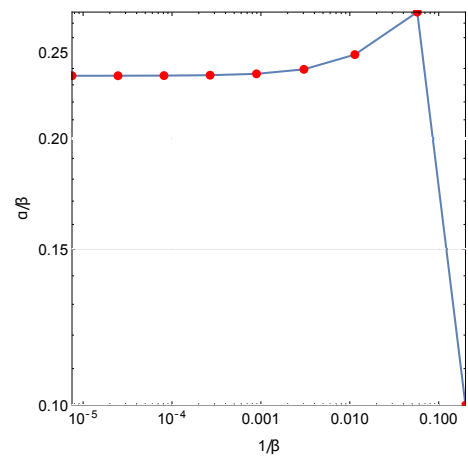, width=6cm}
	\caption{The same FODO cell but with unstable parameters.  It can be seen that the movement of the beam converges to one of the stable points located at the imaginary axis.  Note the logarithmic scale that is required to show all points as the convergence towards the fixed point is exponential.} 
	\label{fig:fodoCell5}
\end{figure}

As a final example, for the sake of comparison, we have increased the strength of the quadrupoles until the Moebius transformation becomes hyperbolic.  In this case, the movement of the particles change dramatically both qualitative and quantitatively, and instead of showing a rotation around one of the fixed points, now the beam trajectory converges to one of the fixed points, which actually is located at the imaginary axis and therefore it represents a beam of infinite size, $\beta \rightarrow \infty$.  This situation is shown in Fig. \ref{fig:fodoCell5}.

\section{Conclusions}
A general formulation of the complex formalism of one dimensional linear beam dynamics has been presented. Through determining and solving the general differential equation governing the beam dynamics in complex form, it has been shown that the general complex transformation of a beam line is a subgroup of the Moebius transformation. Using the proposed formalism, the transformation of several common beam line components has been obtained and graphically represented.  It has been shown that, although the complex formulation is equivalent to the Twiss parameters approach, it generalizes and complements the classical analysis of a beam transport line, it allows to prove some theorems of beam dynamics in a simpler way and it opens the possibility of transforming complete regions of the complex phase space instead of just single points, as is the case in the classical formalism of beam transport lines.

Further, for beam transport through periodic lines, the proposed complex formalism has allowed us to prove the existence of invariant circles under a periodic transformation. In the classical formalism only the invariant points are considered, and they are identified with the actual Twiss parameters for this point location of the transport line. Now, we have seen that the invariant points are nothing but invariant circles of zero radius.  In this way, it is possible to obtain a higher bound of the maximum beam excursion at any point.  Through the classical approach, the mismatched injection on a periodic line is treated as if the Twiss parameters in the injected line are those of the fixed point of the one-period transformation (actually turning an initial conditions problem into an eigenvalue problem), so that the beam will increase its emmitance due to higher order effects to accomodate the new ellipse.  Although this may be an acceptable description for a circular accelerator, where millions of turns under the non-linear field will smear the beam phase space, it may be too pessimistic for a long periodic transfer line.  In this case, the description of the present paper, in which the transformed point moves in the invariant circle may be a better description of  reality.

As a general conclusion, it can be stated that the proposed complex formalism provides a deeper complementary understanding of linear beam dynamics as compared to the classical formalism, it allows to map complete regions of the complex phase plane instead of single points, and in general it contributes to improve our fundamental knowledge of beam transport dynamics.

\section{Acknowledgements}
The authors are grateful to MINECO and UPV/EHU for partial support of this work under grants DPI2017-82373-R and GIU18/196, respectively.

\appendix*

\section{Some useful properties of the Moebius transformation}
The general result given by Eq.~(\ref{eq:moebius1}) can be further exploited by considering the mathematical properties of the Moebius transformations, which have been extensively studied \cite{Needham97},\cite{Campos1},\cite{Campos2}. In this Appendix some results useful for the analysis of beam transfer lines as proposed in this paper are presented and commented.

\subsection{The matrix representation of the Moebius transformation}
A general Moebius transformation, 
\begin{equation}
w=\frac{a z +b}{c z+d}  \textrm{\hspace{1cm}with $ad-bc \ne 0$}
\end{equation}
 
may be represented by a matrix:

\begin{equation}
\mathfrak H =
\begin{bmatrix}
a & b \\
c & d
\end{bmatrix}
\end{equation}

and the composition of Moebius transformations corresponds to the matrix multiplication.  In our case, taking into account Eq.~(\ref{eq:moebius1}), the matrix $\mathfrak H$ will be given by the more restricted form:

\begin{equation}
\mathfrak H =
\begin{bmatrix}
S' & -j C' \\
j S & C
\end{bmatrix}
\end{equation}

which somewhat resembles the transport matrix for a single particle. However, it should be noted here that this case deals with a complex number representing the whole phase space ellipse of the beam.  As $|\mathfrak H|=1$, the Moebius transformation is normalized.

The matrix representation becomes particularly clear if the complex numbers, $w$ and $z$ are expressed by homogeneous coordinates.  In this case, the Moebius transform may be written as:

\begin{equation}
\frac{w_1}{w_2}=\frac{a \frac{z_1}{z_2} +b}{c \frac{z_1}{z_2} +d}=\frac{a z_1+b z_2}{c z_1+ d z_2}
\end{equation}

  Thus, since it is possible to represent a complex number, $z$ by its column vector of homogeneous coordinates, $\underline{z}$, the transformation will be:

\begin{equation}
\underline{w} = \mathfrak{M} \underline{z}
\label{eq:linearTransform}
\end{equation}

\subsection{The fixed points of the transformation}
The fixed points are left invariant by the transformation.  These points can be obtained by imposing the conditon that that the complex point is not changed by the transformation:
\begin{equation}
Y=\frac{a Y+b}{c Y+d}
\end{equation}

Alternatively, the fixed points are defined by the eigenvectors of Eq.~(\ref{eq:linearTransform}).  The solution for the particular case of the beam transport is Eq.~(\ref{eq:fixed}):

\begin{equation}
Y_\pm = \frac{j(C-S') \pm \sqrt{4-(C+S')^2}}{2 S}
\label{eq:fixed}
\end{equation}

There are three possibilities, if $|C+S'| < 2$, Eq.~(\ref{eq:fixed}) will have two complex solutions symmetrical with respect to the imaginary axis, if $|C+S'| = 2$, there will be only one double solution in the imaginary axis and if if $|C+S'| > 2$, there will be two solutions contained in  the imaginary axis.

\subsection{The circle preserving properties}
One interesting possibility that opens when considering a beam line transformation as a complex plane transformation, is that entire regions of the complex plane may be transformed as conformal mappings \cite{Arbarello85}.  For instance, one of the properties of the Moebius transform is that generalized circles are transformed to generalized circles.  It is possible then to find a set of initial conditions, envelope them within a circle and transform the circle along the beam line.  All the initial conditions will remain inside the transformed circle.  Because it is possible to analyze the evolution of the radius of the circle along the transformed planes, it is possible for instance to know if the solutions converge or not and at which speed.

Next we will proceed as in \cite{GeometryOfComplexNumbers}. In order to prove the circle preserving property, first the equation of the circle in the complex plane will be expressed in a more general way.  A circle of radius $\rho$ and center at $\gamma$, may be expressed as:

\begin{equation}
|z-\gamma| = \rho
\end{equation}

or:

\begin{equation}
\left(z-\gamma\right) \overline{\left(z-\gamma\right)}=\rho^2
\end{equation}

\begin{equation}
z \overline{z}-z\overline{\gamma}-\overline{z} \gamma+\left(\gamma \overline{\gamma}-\rho^2\right)=0
\label{eq:circle2}
\end{equation}

Multiplying  Eq.~(\ref{eq:circle2}) by an arbitrary factor $A$, and writing it in matrix form, it results:

\begin{equation}
	\begin{bmatrix}
		\overline{z} & 1
	\end{bmatrix}
	\begin{bmatrix}
		A & -A \gamma\\
		-A \overline{\gamma} & A\left(\gamma \overline{\gamma}-\rho^2\right)
	\end{bmatrix}
	\begin{bmatrix}
		z \\ 1
	\end{bmatrix}=
	\begin{bmatrix}
		\overline{z} & 1
	\end{bmatrix}
	\begin{bmatrix}
		A & B \\
		C & D
	\end{bmatrix}
	\begin{bmatrix}
		z \\ 1
	\end{bmatrix}
	=0
	\label{eq:circle3}
\end{equation}

In this way, the circle will be represented by the Hermitian matrix $\mathfrak{C}=[A B | C D]$, and Eq.~(\ref{eq:circle3}) describing the circle will be compactly expressed as:

\begin{equation}
	\underline{z}^H \mathfrak{C} \underline{z}=0
	\label{eq:circleRepresentation}
\end{equation}

where the superindex $H$ represents the transpose conjugate of a matrix.  It can be seen that by construction  $\mathfrak{C}$ must be Hermitian, as the diagonal elements are real and the non-diagonal elements are conjugate of each other.  The arbitrary factor $A$ has been introduced in order to include the straight lines as a particular case of the circles.  In the projective plane, a line may be considered as a circle with a point at infinite.  With the representation of Eq.~(\ref{eq:circleRepresentation}), all circles and lines of the complex plane may be represented as the quadratic form of a Hermitian matrix with respect to the homogeneous coordinates of the complex plane.

The determinant of the circle matrix $\mathfrak{C}$ is equal to $-A \rho^2$, and it is called the discriminant of the circle.  Real circles will have a negative discriminant.  The discriminant will be zero if $\mathfrak{C}$ represents a line or a zero radius circle.  A positive discriminant is due to a circle of imaginary radius, which cannot be represented in the ordinary complex plane.

In order to check how the Moebius transform changes a given circle, consider a circle defined at the start of the transformation by:

\begin{equation}
	\underline{z}^H \mathfrak{C}_1 \underline{z}=0
\end{equation}

If $\mathfrak{W}$ is the reverse transformation, i.e. the one causing:

\begin{equation}
	\underline{z}=\mathfrak{W} \underline{\omega}
\end{equation}

then the circle will be transformed in the target plane to:

\begin{equation}
	\underline{\omega}^H  \mathfrak{W}^H \mathfrak{C}_1 \mathfrak{W} \underline{\omega}=0
	\label{eq:circleTransformed}
\end{equation}

The matrix inside the quadratic form of Eq.~(\ref{eq:circleTransformed}) is Hermitian as well, and will represent a new circle in the transformed plane.

\begin{equation}
	\mathfrak{C}_2 = \mathfrak{W}^H \mathfrak{C}_1 \mathfrak{W}
\end{equation}

so it is readily established that the Moebius transformation maps circles into circles in the complex plane.

\end{document}